\documentclass[]{spie}

\usepackage[]{graphicx}
\usepackage{amsmath}
\usepackage{mathtools}
\usepackage{color}

\def\xd{\dot{x}}
\def\qm{\frac{q}{m}}
\def\xdd{\ddot{x}}
\def\xddd{\dddot{x}}

\title{Cooling of relativistic electron beams in chirped laser pulses}

\author{Samuel R. Yoffe\supit{a}, Adam Noble\supit{a}, Yevgen
Kravets\supit{b,a}, and Dino A. Jaroszynski\supit{a}
\skiplinehalf
\supit{a}Department of Physics, SUPA, University of Strathclyde,
Glasgow G4 0NG, UK;
\skiplinehalf
\supit{b}Centre de Physique Th\'eorique, \'Ecole Polytechnique, 91120,
Palaiseau, France
}

\authorinfo{Further author information:\\
S.R.Y.: E-mail: sam.yoffe@strath.ac.uk\\
D.A.J.: E-mail: d.a.jaroszynski@strath.ac.uk}

\begin{document} 
\maketitle 

%%%%%%%%%%%%%%%%%%%%%%%%%%%%%%%%%%%%%%%%%%%%%%%%%%%%%%%%%%%%% 
\begin{abstract}

The next few years will see next-generation high-power laser facilities
(such as the Extreme Light Infrastructure) become operational, for which it
is important to understand how interaction with intense laser pulses affects
the bulk properties of a relativistic electron beam. At such high field
intensities, we expect both radiation reaction and quantum effects to play a
significant role in the beam dynamics. The resulting reduction in relative
energy spread (beam cooling) at the expense of mean beam energy predicted by
classical theories of radiation reaction depends only on the energy of the
laser pulse. Quantum effects suppress this cooling, with the dynamics
additionally sensitive to the distribution of energy within the pulse. Since
chirps occur in both the production of high-intensity pulses (CPA) and the
propagation of pulses in media, the effect of using
chirps to modify the pulse shape has been investigated using a
semi-classical extension to the Landau--Lifshitz theory. Results indicate
that even large chirps introduce a significantly smaller change to final
state predictions than going from a classical to quantum model for radiation
reaction, the nature of which can be intuitively understood.
\end{abstract}


%%%%%%%%%%%%%%%%%%%%%%%%%%%%%%%%%%%%%%%%%%%%%%%%%%%%%%%%%%%%%
\section{Introduction}
\label{sec:intro}
In the coming years, a new generation of high-power laser facilities (such
as the Extreme Light Infrastructure (ELI) \cite{url_ELI}) will become operational,
delivering laser intensities well beyond those currently available. The
extremely large fields generated by these pulses will allow qualitatively
new physical regimes to be explored for the first time, in which both
radiation reaction and quantum effects will have a dominant role to play.

An accelerating charge radiates energy (and momentum), and so must
experience a recoil force. The emission must affect the
dynamics of the particle. This is the concept of radiation reaction. Despite
more than a century of investigation, radiation reaction remains a
contentious area of physics. The most widely-accepted classical description of
radiation reaction is based on (independent) attempts to include the effects
of radiation emission by Lorentz \cite{Lorentz1916} and Abraham
\cite{Abraham1932} at the start of the twentieth century, which were later
made fully relativistic by Dirac \cite{Dirac1938}. The
Lorentz--Abraham--Dirac (LAD) equation for a particle of mass $m$ and charge $q$
in an electromagnetic field $F$ reads
\begin{equation}
\label{eq:LAD}
\xdd^a= \frac{f^a_\text{ext}}{m} + \tau \Delta^a{}_b \xddd^b = -\qm F^a{}_b
\xd^b + \tau \big( \xddd^a - \xdd_b \xdd^b \xd^a \big),
\end{equation}
where $f^a_\text{ext} = -q F^a{}_b \xd^b$ is the Lorentz force.
Here, the constant $\tau \coloneqq q^2/6\pi m$ is the `characteristic
time' of the particle
($\simeq6\times 10^{-24}$~s for an electron). An overdot denotes
differentiation with respect to proper time. Indices are raised and lowered
with the metric tensor $\eta=\text{diag}(-1,1,1,1)$, and repeated indices
are summed from 0 to 3.
The $\xd$-orthogonal projection $\Delta^a{}_b \coloneqq
\delta^a_b+\xd^a\xd_b$ removes any $\xd$-component from the \emph{jerk} term
$\xddd$, since $\Delta^a{}_b \xd^b = 0$ due to the normalisation $\xd^a
\xd_a = -1$. This ensures that $\xdd$ is orthogonal to $\xd$, such that the
mass shell condition, $p^a p_a = -m^2$ where $p^a = m \xd^a = (\gamma
m,\vec{p})$, is preserved. Heaviside-Lorentz units are used with $c=1$.

Due to the presence of the third derivative $\xddd$,
the LAD equation exhibits `runaway solutions'. These pathological solutions,
which are highly unphysical and not observed, can be prevented at the cost
of introducing a dependence on all future forces, known as preacceleration.
Despite only forces within the small time
$\tau$ having a significant influence, this response to all future forces
that will act on the particle is not compatible with the concept of
causality. The recent review \cite{Burton2014} describes these problems and
proposed solutions in greater detail.

The moset widely-used alternative to the LAD description was introduced by
Landau and Lifshitz \cite{Landau1962}. The radiation reaction (self-) force
is treated as a perturbation about the Lorentz force,
\begin{equation}
 \xdd^a = -\qm F^a{}_b \xd^b + O(\tau) \qquad \longrightarrow \qquad \xddd^a = -\qm
 \xd^c \partial_c F^a{}_b \xd^b - \qm F^a{}_b \xdd^b + O(\tau),
\end{equation}
where once again the Lorentz force is substituted into the final term on the
right-hand side. Collecting terms to leading order in $\tau$ we arrive at
the Landau--Lifshitz equation:
\begin{align}
 \label{eq:LL}
 \ddot{x}^a = -\frac{q}{m}F^{ab}\dot{x}_b - \tau \frac{q}{m} \left(
 \partial_c F^{ab}\dot{x}_b \dot{x}^c - \frac{q}{m}\Delta^a{}_b F^{bc} F_{cd} \dot{x}^d
\right).
\end{align}
Note that $\Delta^a{}_c F^{cb}\dot{x}_b = F^{ab}\dot{x}_b + \xd^a
F^{cb}\xd_c \xd_b = F^{ab}\dot{x}_b$.

This classical equation has found widespread use amongst the community.
While there is mounting evidence to suggest that equation \eqref{eq:LL} is
valid provided only that quantum effects can be ignored
\cite{Kravets2013a,Spohn2000}, the extreme conditions expected at ELI will
take us into a regime where this is no longer possible. Instead, we must try
to include some of the effects of quantum emission in order to make
predictions for future experimental outcomes.

Quantum effects are typically considered to be negligible if the electric
field observed by the particle $\hat{E}$ is much less than the
Sauter-Schwinger critical field \cite{Sauter1931,Schwinger1951} typical of
QED processes, $E_S = 1.32 \times 10^{18}$ V/m. That is, when the
\emph{quantum nonlinearity parameter} is small,
\begin{equation}
\chi \coloneqq \frac{\hat{E}}{E_S} \ll 1.
\end{equation}
Parameters obtainable at ELI are expected to approach $\chi\sim 0.8$, where
quantum effects cannot be ignored. However, provided paramters are used such
that $\chi^2 \ll 1$ instead, a semi-classical modification to
\eqref{eq:LL} should be valid \cite{Erber1966}.

In order to model the weakly quantum regime, we briefy consider an
important difference between classical and quantum emission. In the
classical case, the particle radiates arbitrarily small amounts of energy at all
frequencies. However, the quantum description requires the particle to emit
entire photons of energy. The frequency of the emitted photons is therefore limited by the energy of the
particle, which introduces a cutoff to the radiation spectrum. The
continuous classical emission which underpins the classical descriptions of
radiation reaction is therefore expected to overestimate the radiation
produced as quantum effects become important \cite{Ritus1979}. The Landau--Lifshitz equation
is therefore extended to include this reduced emission, following Kirk,
Bell and Arka \cite{Kirk2009}, by scaling the radiation reaction force by a
function of the quantum nonlinearity parameter $\chi$. The full expression
for $g(\chi)$ involves a non-trivial integral over Bessel functions of the
second kind. Instead, we use an approximation introduced by Thomas
\textit{et al.} \cite{Thomas2012} based on fitting to numerical data,
\begin{equation}
\label{eq:QM_g}
g\left(\chi\right) = \left(1 + 12\chi + 31\chi^2 + 3.7\chi^3\right)^{-4/9}.
\end{equation}
Predictions using this semi-classical model can be compared to those
obtained from the classical Landau--Lifshitz theory in order to explore the
role of quantum effects in the weakly quantum regime which will be available
at ELI.

%%%%%%%%%%%%%%%%%%%%%%%%%%%%%%%%%%%%%%%%%%%%%%%%%%%%%%%%%%%%%
\section{Collision with intense plane-wave laser pulses}
\label{sec:}

For a linearly polarised plane wave, the electromagnetic field
tensor $F$ depends on spacetime only through the phase $\phi = \omega t -
\vec{k} \cdot \vec{x}$, and takes the form
\begin{equation}
\label{eq:F_planewave}
\frac{q}{m}F^a{}_b = a(\phi) \big(\epsilon^a k_b - k^a \epsilon_b
\big),
\end{equation}
where $k = (\omega,\vec{k})$ is the (null) propagation direction of the laser and
$\epsilon$ the othogonal (transverse) polarisation vector. The function
$a(\phi)$ is a dimensionless measure of the electric field strength. An
$N$-cycle pulse modulated by a $\sin^2$-envelope has been used, for which
\begin{equation}
 \label{eq:pulse}
 a(\phi) = \left\{
   \begin{array}{ll}
     a_0 \sin(\phi) \sin^2\left(\pi \phi / L \right) & \text{for }0 < \phi
     < L \\
     0 & \text{otherwise}
   \end{array} \right. ,
\end{equation}
where $a_0$ is the dimensionless (peak) intensity parameter (or
\emph{normalised vector potential}) and $L = 2\pi N$ is the (total) pulse
length (full-width half-maximum duration is $L/2$). This pulse shape
offers compact support, allowing the particles to begin and end in vacuum.

The total fluence (energy per unit area) of the pulse is proportional to
\begin{equation}
 \mathcal{E} = \int_0^L d\phi\ a^2(\phi) = \frac{3\pi}{8} N a_0^2.
\end{equation}
In this work, $\mathcal{E}$ is kept constant, which fixes $a_0$ for each
$N$. The classical Landau--Lifshitz prediction for the final-state particle
distribution is completely determined by the
fluence\cite{Neitz2014,Yoffe2015}. By contrast, quantum effects are expected to
depend directly on the intensity of the field itself, that is on the value
of $a_0$. Maintaining constant fluence has the advantage that quantum
effects can be studied while maintaining the same classical prediction.

Since our laser pulse is described by a plane wave, we consider the initial
momenta to be strongly peaked around zero in the transverse
directions\cite{Yoffe2015} and focus on the longitudinal properties of the
distribution. The initial electron beam is then taken to be a Maxwellian
distribution for the longitudinal momentum $p$ (in units of $mc$):
\begin{equation}
 \label{eq:initial_dist}
 f\left(\phi=0,p\right) = \frac{N_P}{\sqrt{2\pi\theta}}
 \exp\left[-\frac{(p-\bar{p})^2}{2\theta}\right],
\end{equation}
where $\theta$ is the variance of the distribution about its mean $\bar{p}$,
and $N_P$ is the number of particles used to represent it. Instead of randomly
sampling the distribution using a large number of particles, an iterative
technique is used which facilitates efficient and accurate reconstruction of
the distribution using far fewer particles\cite{Yoffe2015}. For the case of
classical radiation reaction according to the Landau--Lifshitz equation,
this method has been shown to be in excellent
agreement with the analytical solution of the Vlasov
equation\cite{Yoffe2015,Noble2013a}.

In order to describe the properties of the distribution, we introduce the
\emph{relative momentum spread} and the \emph{momentum skewness},
\begin{align}
 \label{eq:properties}
 \hat{\sigma}(\phi) = \frac{\sqrt{\theta(\phi)}}{\bar{p}(\phi)} \quad
 \text{and} \quad S(\phi) = \frac{\left\langle \big[p-\bar{p}(\phi)\big]^3
 \right\rangle}{\theta^{3/2}(\phi)},
\end{align}
respectively. The former gives a measure of the beam quality, while the latter indicates
how symmetric the distribution is about its mean.

\begin{figure}[!tb]
 \begin{center}
  \includegraphics[width=0.47\textwidth]{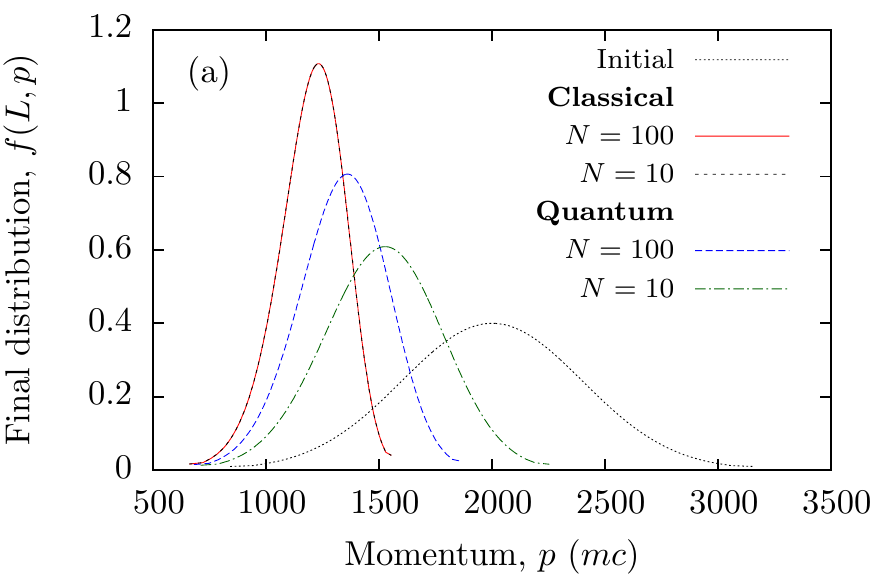}
  \includegraphics[width=0.47\textwidth]{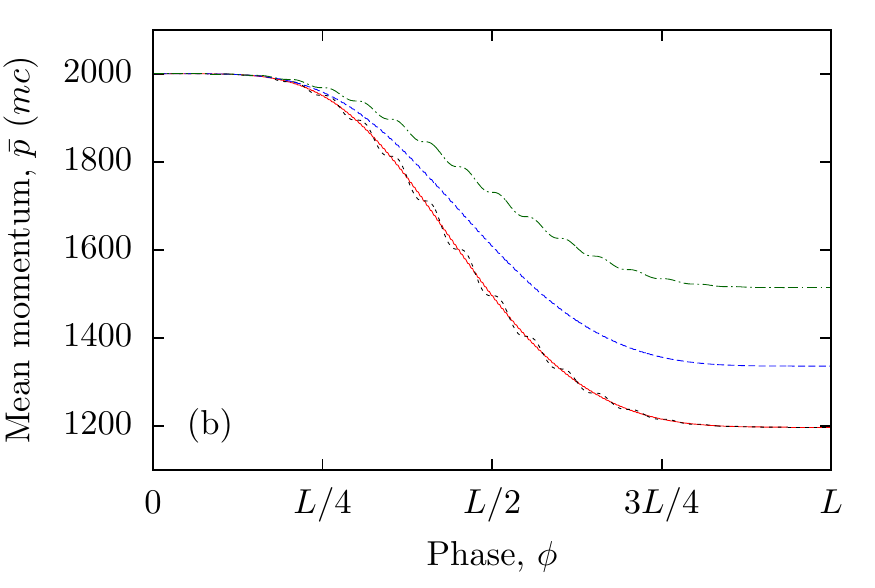} \\
  \includegraphics[width=0.47\textwidth]{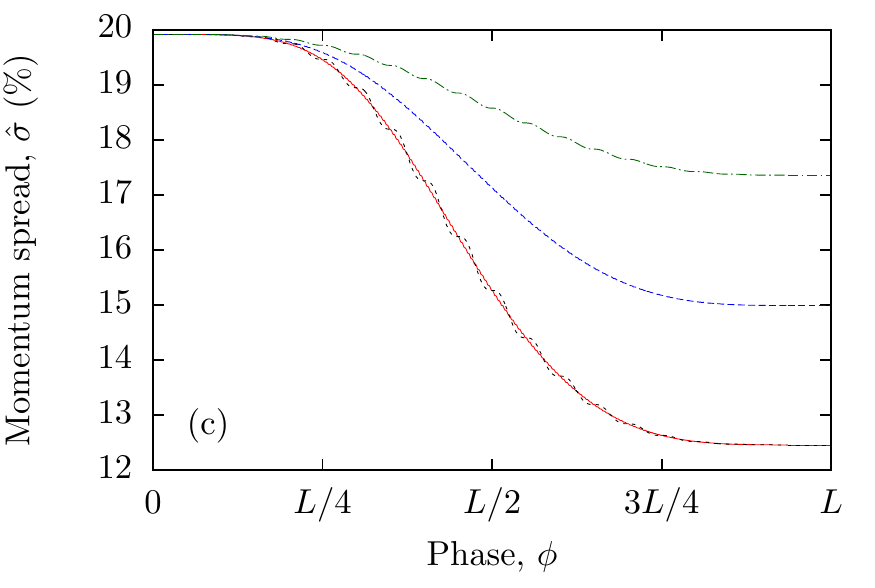}
  \includegraphics[width=0.47\textwidth]{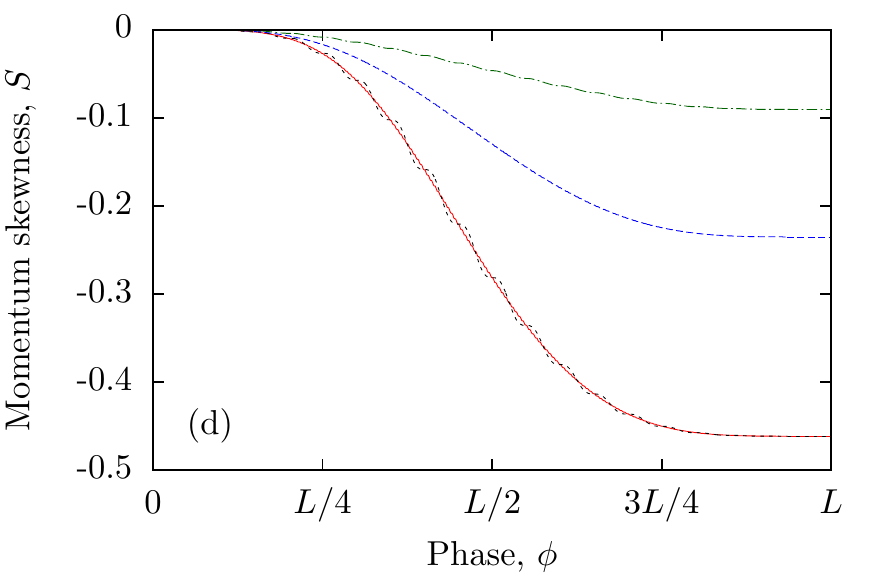}
 \end{center}
 \caption{(Colour online) Comparison of classical and semi-classical
 predictions for the final-state properties of an electron beam colliding
 with pulses of $N = 10,100$ cycles. Part (a): initial and final electron
 distributions. Classical predictions for both $N = 10$ and 100 lie on top
 of one another. Parts (b)--(d): Evolution of the mean momentum $\bar{p}$,
 relative momentum spread $\hat{\sigma}$, and mometum skewness during the
 interaction. Classical results for the final distribution show beam cooling
 with a dependence on the total fluence only, while the quantum model
 predicts reduced cooling dependent on the pulse intesnsity.}
 \label{fig:results}
\end{figure}

We now consider the interaction of an energetic electron beam with a
high-intensity laser pulse. Parameters have been chosen to be within the
capabilities of ELI. An electron beam represented by $N_P = 401$ particles
with an initial momentum spread of 20\% around $\sqrt{1 + \bar{p}^2} = 2
\times 10^{3}$ (which corresponds to an energy of approximately 1 GeV) is collided
with pulses of length $N \in [5,200]$ cycles with constant $N a_0^2 =
9248$. For $N = 20$ with a wavelength of $\lambda = 800$~nm, this
corresponds to a full-width half-maximum pulse duration of $27$~fs with peak
intensity $2\times10^{21}$~W/cm$^2$.

Figure \ref{fig:results} presents a comparison of classical and semi-classical
predictions for the cooling of a relativistic electron beam in the
interaction with two pulse lengths, $N = 10$ and $N = 100$. For a wavelength of
$\lambda = 800$ nm, the former represents a pulse of total duration 27 fs
with peak intensity $I = 3.96 \times 10^{21}$ W/cm$^2$, while the latter a
total duration of 270 fs and $I = 3.96 \times 10^{20}$ W/cm$^2$. These
intensities should be well within the capabilities at ELI \emph{without the
need for extreme focussing}. These unfocused pulses are better described by
a plane wave.

Figure \ref{fig:results}(a) shows the initial and final particle distribution
for both pulse lengths predicted by both classical and semi-classical
models. It should be noted that the classical final distributions sit
directly on top of one another, supporting the statement that the
final state classical predictions depend only on the total fluence of the
laser, and not how the energy is distributed within it. The final
distribution is significantly more sharply peaked around a much lower
momentum than the initial distribution, demonstrating beam cooling.
It also shows signs of antisymmetry, with a longer tail extending
to lower momenta.
In contrast, the semi-classical model predicts a different final
distribution for each $N$. For $N = 100$, the dimensionless intensity
parameter $a_0 \simeq 9.6$ and we see that quantum effects have become
important. Beam cooling has been reduced, and this trend continues to $N =
10$ (where $a_0 \simeq 30$).

The fluctuation of the mean momentum as the beam passes through the pulse is
shown in Fig. \ref{fig:results}(b), with the relative momentum spread
presented in Fig. \ref{fig:results}(c). For the classical cases, we see
that, despite the different pulse lengths behaving slightly differently
during the evolution due to the different numbers of cycles, both converge
on the same final state with $\bar{p} = 1197.70$ and $\hat{\sigma} = 12.46\%$.
For comparison, the analytical solution of the Vlasov equation with
the Landau--Lifshitz theory for radiation reaction\cite{Yoffe2015} predicts
$\bar{p} = 1197.62$ and $\hat{\sigma} = 12.53\%$. Instead, the
semi-classical model for $N = 100$ predicts $\bar{p} = 1336.28$ and
$\hat{\sigma} = 14.99\%$. The results for $\hat{\sigma}$ show a reduction in
beam cooling\footnote{Reduction in beam cooling means that there is
\emph{less} contraction of phase space than would be observed in the
classical case. As such, the relative momentum spread remains larger than
that predicted by the Landau--Lifshitz theory.} ($\hat{\sigma}$ does not reduce as much from the initial 20\%)
with the beam remaining more energetic. For $N = 10$, the semi-classical
model predicts $\bar{p} = 1514.40$ and $\hat{\sigma} = 17.36\%$, continuing
the reduction of beam cooling. Note that, for $N = 100$, we measure the
average quantum nonlinear parameter $\langle \chi \rangle^2 < \langle \chi^2
\rangle < 1 \times 10^{-4}$. For the more intense pulse with $N = 10$, this
increases to $\langle \chi^2 \rangle < 0.12$ and we expect the
semi-classical model to remain a valid quantum approximation.

The momentum skewness of the distribution is given in Fig.
\ref{fig:results}(d), which shows the initially Gaussian ($S~=~0$) profile
becoming increasingly negatively skewed as the evolution proceeds. The
analytical solution of the classical Vlasov equation\cite{Yoffe2015}
predicts a final skewness of $S = -0.498$, while the numerical approach used
here gives $S = -0.462$\footnote{This numerical difference is due to the
finite number of particles used to represent the
distribution\cite{Yoffe2015}.}. This is reflected in Fig.
\ref{fig:results}(a) by the extended tail of the classical final
distribution to lower momenta. This is a consequence of classical beam
cooling, since it is the more energetic particles which are most influenced
by radiation reaction. The semi-classical model suppresses radiation
reaction more for precisely those particles, and as such the distribution
becomes less negatively skewed.

%%%%%%%%%%%%%%%%%%%%%%%%%%%%%%%%%%%%%%%%%%%%%%%%%%%%%%%%%%%%%
\section{Interaction with chirped pulses}
\label{sec:chirps}

The use of chirped-pulse amplification (CPA) to achieve
high-intensity pulses makes their influence on beam dynamics directly relevant to future experiments. In
addition, chirps naturally arise in the propagation of laser pulses in
media. The results of the previous section indicate that, unlike the classical
Landau--Lifshitz theory of radiation reaction, the semi-classical model is
sensitive to \emph{how the energy is distributed} in the pulse, not just the
total energy. To investigate this further, we include a chirp in our plane
wave model for the laser pulse.

We introduce the pulse length for an $N$-cycle chirped pulse
\begin{equation}
 \label{eq:pulse_length}
 L_\Delta = \frac{2\pi N}{1 + \Delta/2},
\end{equation}
where $\Delta$ is the chirp rate. The linearly-chirped phase is then defined
to be
\begin{equation}
 \eta(\phi;\Delta) = \phi \left( 1 + \frac{\phi \Delta}{2L_\Delta}
 \right).
\end{equation}
If $\omega = \partial\phi/\partial t$ is our unchirped frequency, then we
find
\begin{align}
 \Omega(\phi;\Delta) = \frac{\partial}{\partial t}\eta(\phi;\Delta) 
 = \omega \left(1 + \frac{\phi\Delta}{L_\Delta} \right),
\end{align}
indicating that the phase contains a linear chirp.
Thus, at the end of the pulse $\phi = L_\Delta$ the frequency is
$\Omega(L;\Delta) = \omega(1 + \Delta)$. (A \emph{positive} chirp $\Delta > 0$
therefore exhibits an \emph{increase} in frequency as we move through the pulse.)
\begin{figure}[!tb]
 \centering
 \includegraphics[width=0.94\textwidth]{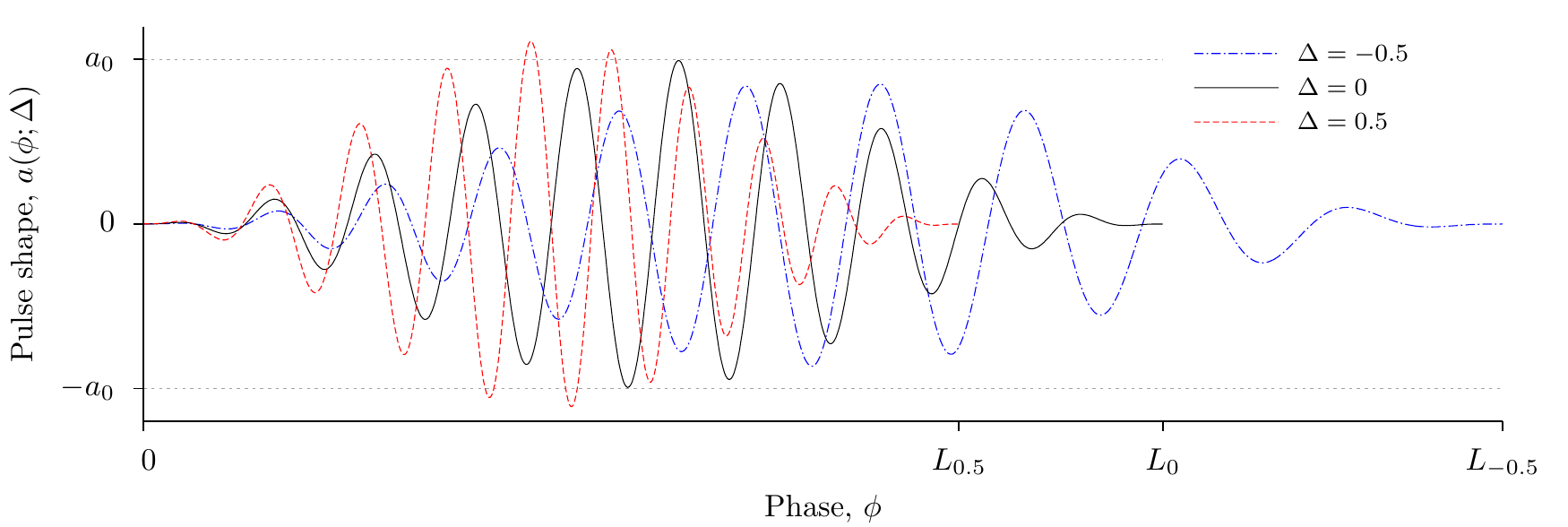}
 \caption{(Colour online) The dimensionless intensity profile for $N = 10$
 pulses with chirp rate $\Delta = \pm 0.5$, along with the original
 unchirped case. The pulse length $L_\Delta$ is
 given by equation \eqref{eq:pulse_length}.}
 \label{fig:chirp_pulse}
\end{figure}
The description of the pulse given
in equation~\eqref{eq:pulse} is then replaced by
\begin{equation}
 \label{eq:chirped_pulse}
 a(\phi;\Delta) = \left\{
   \begin{array}{ll}
    a_0 \sqrt{1 + \frac{\Delta}{2}} \sin \Big[ \eta(\phi;\Delta) \Big]
 \sin^2 \left(\frac{\pi \phi}{L_\Delta} \right) & \text{for}\quad 0 < \phi
 < L_\Delta \\
    0 & \text{otherwise}
   \end{array} \right. .
\end{equation}
The electron beam encounters the same frequency at the front of the pulse
for both positive and negative chirps, with the observed frequency then
increasing (positive chirp) or decreasing (negative chirp) towards the
rear of the pulse. The inclusion of the factor $\sqrt{1 + \Delta/2} =
\sqrt{L_0/L_\Delta}$ in equation~\eqref{eq:chirped_pulse} ensures that
pulses with the same $N$ and $a_0$ contain the same fluence, regardless of
the chirp rate. This is important, as we want to investigate the role of the
chirp itself. A comparison of pulses of different energy, not just
different distribution of energy, would obscure this.

The pulse shape $a(\phi;\Delta)$ is shown in Fig.
\ref{fig:chirp_pulse} for $\Delta = \pm 0.5$, along with the unchirped
pulse (\textbf{------}). This illustrates how the increase in frequency caused by the positive
chirp ({\color{red}\textbf{-- -- --}}) generates a shorter pulse length than
the unchirped pulse. The amplitude is increased from $a_0$ (indicated by the
horizontal dotted lines) such that the
fluence remains constant. The negative chirp ({\color{blue}\textbf{--
$\cdot$ --}}), on the other hand, causes an
extended pulse of lower peak intensity for the same fluence.

\begin{figure}[!tb]
 \centering
 \includegraphics[width=0.47\textwidth]{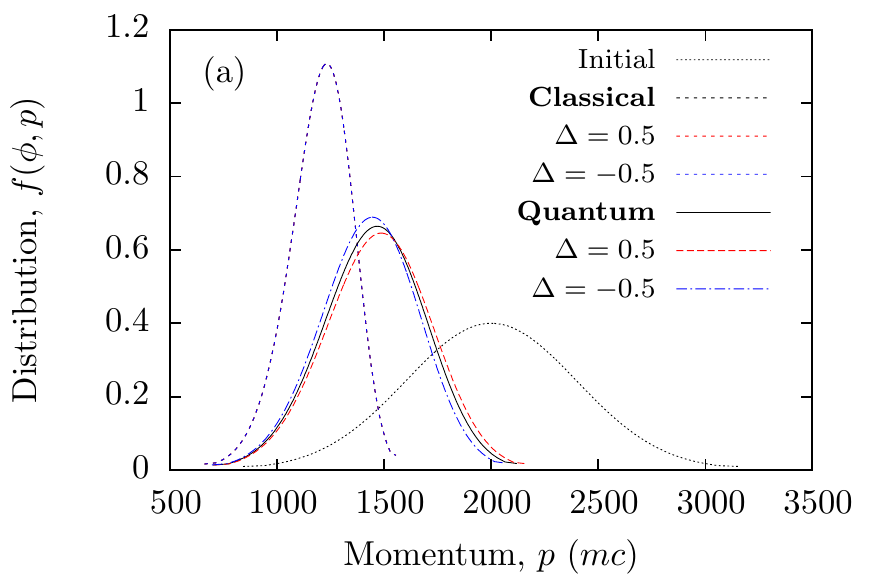}
 \includegraphics[width=0.47\textwidth]{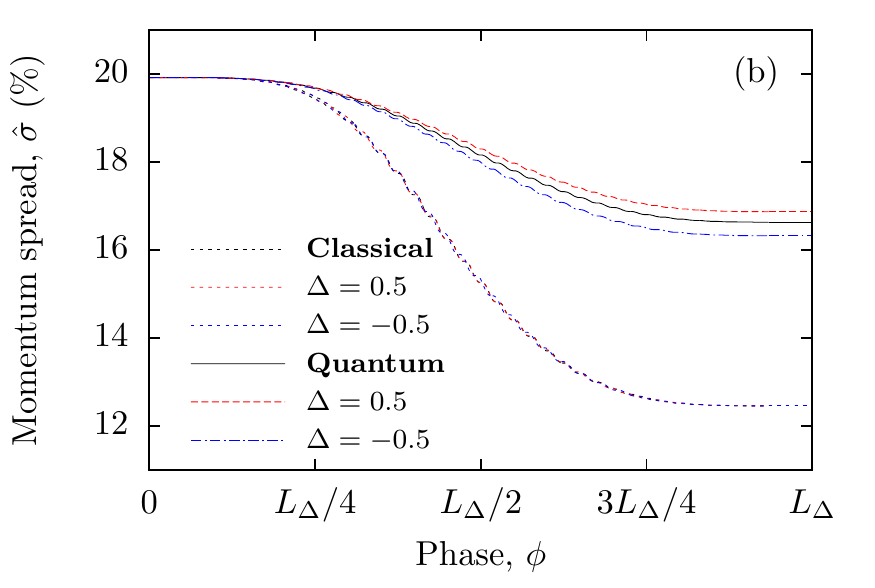}
 \caption{(Colour online) The influence of chirping in the semi-classical
 model of radiation reaction. Comparison of (a) initial and final
 distributions, and (b) evolution of the spread $\hat{\sigma}$, with and
 without a chirp. The classical final distribution and spread are
 insensitive to the chirp, despite a small variation during the evolution.}
 \label{fig:chirps}
\end{figure}

Since the fluence is kept constant, predictions from the classical
Landau--Lifshitz theory for the final properties of the beam are not
affected by a chirp. This is confirmed in Fig.~\ref{fig:chirps}, where the
influence of a chirp in the semi-classical model is also shown for the
interaction of a 1 GeV electron beam with $N = 20$ ($a_0 \simeq 22$)
laser pulses.
Figure~\ref{fig:chirps}(a) shows the initial and final distributions
obtained using both the classical and semi-classical models, with and
without chirps with $\Delta = \pm0.5$. The evolution of the spread
$\hat{\sigma}$ as the bunch travels through the pulse is shown in
Fig.~\ref{fig:chirps}(b), which highlights the significant \emph{reduction
in beam cooling} when
using the semi-classical model. Even with a relatively large chirp of
$\lvert\Delta\rvert = 0.5$, the impact of chirping the laser pulse can been
seen to be a notably smaller effect than moving from the classical to the
semi-classical model.

For the positively chirped case, a final spread $\hat{\sigma}_+ = 16.9\%$
is found, compared to $\hat{\sigma}_0 = 16.6\%$ for the unchirped pulse and
$\hat{\sigma}_- = 16.3\%$ for the negative chirp.
These changes in the spread are easily understood by considering the peak
intensity encountered in each case. For the positive chirp, the pulse length
is reduced because of the decrease in wavelength as we move through the
pulse. To keep the fluence the same as for the unchirped pulse, the peak
intensity must increase from $a_0$ to $a_0\sqrt{1 + \Delta/2}$. In turn,
this increase leads to a higher instantaneous value of the quantum
parameter, $\chi$, and thus increased suppression of the radiation reaction
effect. Hence, beam cooling is further reduced by a positive chirp. In
contrast, a negatively chirped pulse has a longer duration and a
corresponding lower peak intensity, therefore quantum effects are less
important and the radiation reaction term is less strongly suppressed.

\section{Conclusions}
In the next few years, as new high-powered laser facilities come online, it
is increasingly important to understand the physics that will occur at these
unprecedented field strengths. Previously untested areas of physics will be
experimentally probed for the first time, which requires a knowledge of
fundamental principles such as radiation reaction in regimes for which
quantum effects can no longer be ignored.

In this paper, a semi-classical extension to the Landau--Lifshitz theory of
radiation reaction has been used to investigate the role of chirps in the
interaction of relativistic electron beams with high-intensity laser pulses.
Comparison is first made to classical predictions to identify the
differences as some quantum effects are included in the model. Classical
results are compared to analytical solution of the Vlasov equation,
supporting the validity of the numerical technique used. A reduction
in the amount of beam cooling is observed in the quantum case and
found to depend on the intensity profile of the pulse itself, unlike
classical predictions which only care about the total energy of the pulse.
Chirps have been used to alter the distribution of energy within the
pulses, with even a large chirp of 50\% resulting in only a small
modification to the final state properties, the origin of which is
intuitively understood.

The results presented here are limited to the semi-classical case
$\chi^2 \ll 1$ due to the deterministic equation of motion used to describe
radiation reaction. Strongly quantum regimes may be explored using similar
techniques by adopting a stochastic equation where photon emission
probabilities are determined by strong field QED, as in
\cite{Elkina2011,Green2014}. This will be addressed in future work.

\acknowledgments
This work is supported by the UK EPSRC (Grant EP/J018171/1); the ELI-NP
Project; and the European Commission FP7 projects Laserlab-Europe (Grant
284464) and EuCARD-2 (Grant 312453). Datasets available online
\cite{Yoffe2015_D1,Yoffe2015_D2}.

%%%%%%%%%%%%%%%%%%%%%%%%%%%%%%%%%%%%%%%%%%%%%%%%%%%%%%%%%%%%%
%%%%% References %%%%%

\bibliography{refs}
\bibliographystyle{spiebib}

\end{document}